# (Accepted Manuscript) Comparison of global sensitivity analysis methods for a fire spread model with a segmented characteristic

Shi-Shun Chen, Xiao-Yang Li*

School of Reliability and Systems Engineering, Beihang University, Beijing 100191, China

Email: css1107@buaa.edu.cn (Shi-Shun Chen), leexy@buaa.edu.cn (Xiao-Yang Li)

Corresponding author*: Xiao-Yang Li

**This manuscript has been accepted by *Mathematics and Computers in Simulation***

Article link: https://www.sciencedirect.com/science/article/pii/S0378475424004014

Code link: https://github.com/dirge1/GSA_segmented

## Highlights

- The effect of segmented characteristics on GSA is explored by a fire spread model.
- Four GSA methods give different importance rankings during the transition region.
- The Sobol index yields a radical importance ranking.
- Analysts should choose GSA methods carefully according to their practical purpose.

## Abstract

Global sensitivity analysis (GSA) can provide rich information for controlling output uncertainty. In practical applications, segmented models are commonly used to describe an abrupt model change. For segmented models, the complicated uncertainty propagation during the transition region may lead to different importance rankings of different GSA methods. If an unsuitable GSA method is applied, misleading results will be obtained, resulting in suboptimal or even wrong decisions. In this paper, four GSA indices, i.e., Sobol index, mutual information, delta index and PAWN index, are applied for a segmented fire spread model (Dry Eucalypt). The results show that four GSA indices give different importance rankings during the transition region since segmented characteristics affect different GSA indices in different ways. We suggest that analysts should rely on the results of different GSA indices according to their practical purpose, especially when making decisions for segmented models during the transition region. All of our source codes are publicly available at https://github.com/dirge1/GSA_segmented.

# 1 Introduction

Computational models are always adopted to describe real systems. However, due to data sparsity, environmental noise or lack of knowledge, uncertainties are always involved in model inputs, resulting in uncertain model outputs [1]. At this point, global sensitivity analysis (GSA) is a powerful tool to assess the impact of the uncertainty in each input variable on the uncertainty of model output [2]. With the help of GSA, analysts can find out effective ways of model calibration, uncertainty reduction and risk control [3-5]. As a result, GSA has received extensive attention in practical applications, such as pandemic progression [6], environmental dynamics [7], and engineering design [8, 9].

Among all the GSA indices in the literature, variance-based methods are the most widely used and well-proven approaches [10-12]. Nevertheless, they rely on the assumption that variance is sufficient to describe the output uncertainty. Since variance-based methods only consider the second-order moment, some information is inevitably lost when output uncertainty is solely represented with variance [13]. To overcome this limitation, numerous moment-independent GSA methods have been proposed. The first representative moment-independent method is the delta index developed by Borgonovo [14]. The delta index is a normalized index that quantifies the sensitivity based on the area difference between the conditional and marginal probability density function (PDF) of output. The mutual information derived from information theory is an alternative moment-independent GSA approach [15], which can quantify the reduction in the uncertainty of the target variable (i.e., model output) when controlling the uncertainty of a source variable (i.e., model input) based on the conditional and marginal PDF of output. Further, Pianosi and Wagener [16] proposed another moment-independent GSA measure based on the conditional and marginal cumulative distribution function (CDF) of output, called PAWN index, to avoid the estimation of PDFs, thus improving the robustness of the index and cutting down the computation time. For detailed introduction of other GSA approaches, interested readers are referred to [17].

Due to the existence of numerous GSA methods, the pros and cons of different GSA methods are of great interest. In this regard, scholars have conducted many comparative studies. For instance, Ikonen [18] and Li et al. [19] compared the variance-based method, delta index and the elementary effects method by application to fuel behavior modeling and thermal hydraulic phenomena, respectively. Iooss and Prieur [20] and Vuillod et al. [21] compared the Shapley effect with the variance-based method for models with independent and correlated inputs, respectively. They found that different GSA methods had similar results for variables with higher importance. Rabitti and Borgonovo [22] compared the delta index with the Shapley effect for a mathematical annuity model. They demonstrated that both methods could give reliable answers in complicated case with correlated inputs. Upreti et al. [23] and Khorashadi et al. [24] compared the PAWN index with the variance-based method for the Aquacrop model and SWAT model, respectively. They concluded that when the output PDF was highly-skewed, both methods could identify the non-influential variables, and the PAWN index could quantify the relative importance of the influential variables more effectively than the variance-based methods.

In the above studies, scholars mainly focus on the computational efficiency and convergence of the GSA indices and different GSA methods get similar importance rankings. Whereas, in some cases, the results obtained by different GSA methods can be quite different. For example, Borgonovo [25] compared the delta index with the variance-based method for a risk assessment model. He revealed that the influential variables identified by the variance-based methods were not always consistent with the ones that influenced the entire output distribution the most. Therefore, applying moment-independent methods, like the delta index, was necessary for decision making. In [26], the authors compared the entropy-based importance measure with the variance-based method, mutual information and delta index for the Ishigami function. They found that different GSA methods get different importance rankings due to the undesirable characteristics of the output PDF.

The above research shows that different GSA methods may get different importance rankings. However, in practical applications, most analysts tend to use the GSA method that they are comfortable with, rather than the method that best suits the purpose and problem at hand [17]. If different GSA indices give different importance rankings for the target model and an unsuitable GSA index is applied, misleading results will be obtained, which will lead to suboptimal or even wrong decisions. Therefore, it is crucial for analysts to know the applicability of different GSA methods to their target model.

In this study, we focus on the implementation of GSA on a special class of mathematical models, i.e. segmented models. Segmented models are commonly adopted to describe the behavior of some special systems whose input-output relationship may be changed when certain indicator exceeds the corresponding critical threshold. For example, many scholars suggested that with the increase in wind speed, the influence of atmospheric conditions on fire spread would change [27, 28]. As a result, the critical threshold of wind speed should be identified and different models need to be developed for different stages.

For segmented models, due to the uncertainty in the indicator or threshold, there is a transition region where the behavior of the system cannot be accurately determined. During the transition region, the complicated uncertainty propagation may lead to some undesirable characteristics in the output PDF (e.g., highly-skewed or heavily-tailed). Hence, it is important to discuss and compare the results of different GSA methods on the segmented model critically, so as to provide credible information for decision making. Indeed, some researchers have conducted GSA for segmented models. However, most of the relevant studies only utilized the variance-based method. For instance, Spiessl and Becker [29] compared three computationally efficient variance-based methods and three sampling schemes by application to a final repository model with quasi-discrete behavior. Liang et al. [30] conducted GSA for four representative nitrogen uptake models of crops by the variance-based method. In particular, the DAISY model considered in their research was a three-stage segmented model, in which the nitrogen uptake corresponded to different models at different accumulated temperature levels. Lamboni et al. [31] employed the variance-based method to perform GSA for a dynamic crop model, where the denitrification process was a two-stage segmented model determined by temperature. In [32], the authors developed a simple model for the establishment of tick-borne pathogens where the growth mechanism of larvae and nymphs was changed at different periods, corresponding to different models. Then, they carried out GSA for the model by the variance-based approach.

With the development of the moment-independent methods, recently, Kc et al. [33] analyzed a segmented fire spread model using the PAWN index and compared the GSA results with the variance-based ones. Their results illustrated that the PAWN index should be prioritized when computational resources are limited, while the variance-based approach was able to provide more detailed information about uncertainty propagation. Nonetheless, their study only focused on the second stage of the fire spread model rather than the transition region between the two stages, which means that they did not consider the impact of segmented characteristics on the GSA results. Table 1 provides a summary of the above literature reviews. As a result, to the best of our knowledge, none of the existing studies discuss and compare the results of different GSA methods on segmented models, especially in the transition region.

Table 1 Summary of the literature reviews.

| Literature | Sobol index | Delta index | Mutual information | PAWN index | Focus on segmented models | Consider the transition region |
|---|---|---|---|---|---|---|
| Ikonen [18] | √ | √ | | | | |
| Li et al. [19] | √ | √ | | | | |
| Upreti et al. [23] | √ | | | √ | | |
| Khorashadi et al. [24] | √ | | | √ | | |
| Borgonovo [25] | √ | √ | | | | |
| Tang et al. [26] | √ | | √ | | | |
| Spiessl and Becker [29] | √ | | | | √ | |
| Liang et al. [30] | √ | | | | √ | |
| Lamboni et al. [31] | √ | | | | √ | |
| Dunn et al. [32] | √ | | | | √ | |
| Kc et al. [33] | √ | | | √ | √ | |
| This work | √ | √ | √ | √ | √ | √ |

Focusing on this critical issue, in this paper, four GSA indices, i.e. the Sobol index, mutual information, delta index and PAWN index, are employed for the aforementioned segmented fire spread model introduced by Kc et al. [33]. It is an empirical fire spread model established by Cheney et al [34] called Dry Eucalypt model, which is widely used for Australian eucalypt forests. We primarily concentrate on the importance rankings in the transition region of the segmented model. Then, as can be seen in Section 3, we find that the results of these four GSA methods are not the same during the transition region. Motivated by this phenomenon, a specific case is chosen and the reasons for the inconsistent results are further analyzed by fixing variables to different values in turn and comparing conditional PDF and CDF to the original ones. Furthermore, we conduct comparisons between the GSA methods and discuss how to choose appropriate GSA methods. The main aim of this paper is to demonstrate the specificity of conducting GSA on segmented models and guide the choice of GSA methods.

## 2 Materials and methods

### 2.1 GSA methods

In order to facilitate the description of subsequent GSA methods, the following generic model is introduced. Assume that the output response of a system is determined by a multitude of input random variables, which can be expressed as

$$Y = g(\boldsymbol{X}), \tag{1}$$

where $\boldsymbol{X} = (X_1, X_2, \cdots, X_n)$ is an $n$-dimensional vector of the input variables with random uncertainties; $Y$ represents the output response with uncertainty propagated by $\boldsymbol{X}$ via the function $g(\cdot)$. It should be noted that the inputs and output represented in Eq. (1) are random variables. The goal of GSA is to quantify the uncertainty importance of each input variable in respect to the output variable from their samples. In the following subsections, we will introduce the GSA methods based on Eq. (1). The numerical implementations of these four GSA indices are presented in the Appendix.

2.1.1 Sobol index

The Sobol index is based on the total variance decomposition given by [35]

$$V(Y) = \sum_{i=1}^{n} V_i + \sum_{i=1}^{n-1} \sum_{j=i+1}^{n} V_{i,j} + \cdots + V_{1,2,\cdots,n}, \tag{2}$$

where $V(\cdot)$ stands for the variance; $V_i$ is the variance contributed by the input variable $X_i$ to the output variance individually; $V_{i,j}$ is the variance contributed by the interaction between variable $X_i$ and $X_j$ and $V_{1,2,\ldots,n}$ is the variance caused by the interaction among all variables.

According to [35], the total variance of $X_i$ considering its own variance and all its interactions with other variables can be derived as

$$V_i^T = V_i + \sum_{j \neq i}^{n} V_{ij} + V_{1,2,\cdots,k} = V(Y) - V\left[E\left(Y \middle| \boldsymbol{X}_{\sim i}\right)\right], \tag{3}$$

where $E(\cdot)$ stands for the expectation; $\boldsymbol{X}_{\sim i}$ denotes all the input variables except $X_i$. Then, two normalized sensitivity indices for $X_i$ can be defined as [35]

$$S_i = \frac{V_i}{V(Y)} = \frac{V\left[\left(E \middle| X_i\right)\right]}{V(Y)} \tag{4}$$

$$S_i^T = \frac{V_i^T}{V(Y)} = \frac{E\left[V\left(Y \middle| \boldsymbol{X}_{\sim i}\right)\right]}{V(Y)} = 1 - \frac{V\left[E\left(Y \middle| \boldsymbol{X}_{\sim i}\right)\right]}{V(Y)}, \tag{5}$$

where $S_i$ and $S_i^T$ represents the main effect and the total effect of $X_i$ on $Y$, respectively.

2.1.2 Mutual information

Shannon defined the differential entropy of continuous variables as [15]

$$H(Y) = -\int_{\Omega_Y} f_Y(y) \ln\left[f_Y(y)\right] \mathrm{d}y, \tag{6}$$

where $\Omega_Y$ is the range of variation of $Y$; and $f_Y(y)$ is the marginal PDF of $Y$. The mutual information $I(X_i; Y)$ between two continuous random variables $X_i$ and $Y$ is defined by

$$I(X_i;Y) = H(Y) - H\left(Y \middle| X_i\right) = \int_{\Omega_{Xi}} \int_{\Omega_Y} f(x_i, y) \ln\left[\frac{f(x_i, y)}{f_{X_i}(x_i) f_Y(y)}\right] \mathrm{d}y \mathrm{d}x_i, \tag{7}$$

where

$$H\left(Y|X_i\right)=-\int_{\Omega_{Xi}}\int_{\Omega_Y} f\left(x_i,y\right)\ln\left[f\left(y|x_i\right)\right]\mathrm{d}y\mathrm{d}x_i \tag{8}$$

is the conditional entropy of $Y$ given knowledge of $X_i$; $f(x_i, y)$ is the joint PDF of $X_i$ and $Y$; $f(y|x_i)$ is the conditional PDF of $y$ given $x_i$; $f_{X_i}(x_i)$ is the marginal PDF of $X_i$; $\Omega_{Xi}$ is the range of variation of $X_i$.

For continuous variables, there is a normalized version of mutual information given by [36]

$$\rho_i=\left[1-\exp\left(-2I\left(X_i;Y\right)\right)\right]^{1/2}. \tag{9}$$

However, when the mutual information is large, this normalized approach may invisible the change in the index. Therefore, the mutual information given by Eq. (7) is straightforwardly considered as a sensitivity index in this paper, denoted as $\eta_i$.

#### 2.1.3 Delta index

To investigate the impact of the input variable uncertainties on the PDF of model output, Borgonovo proposed the delta index as [14]

$$\delta_i=\frac{1}{2}E_{X_i}\left[s\left(X_i\right)\right]=\frac{1}{2}\int f_{X_i}\left(x_i\right)s\left(X_i\right)\mathrm{d}x_i \tag{10}$$

where $s(X_i)$ is the area difference between the marginal output PDF $f_Y(y)$ and the conditional one $f(y|x_i)$, i.e.

$$s\left(X_i\right)=\int_{-\infty}^{+\infty}\left|f_Y\left(y\right)-f\left(y|x_i\right)\right|\mathrm{d}y \tag{11}$$

#### 2.1.4 PAWN index

Unlike the aforementioned moment-independent approaches, the PAWN index uses the cumulative distribution function (CDF) of the output to quantify the sensitivity of input variables. The PAWN index was defined as [16]

$$\kappa_i=\underset{x_i}{\mathrm{stat}}\max_{y}\left|F_Y\left(y\right)-F\left(y|x_i\right)\right| \tag{12}$$

where $F_Y(y)$ and $F(y|x_i)$ are the unconditional and conditional CDF of $y$; and stat is a statistic operator (e.g. maximum, median or mean) defined by the user. In this paper, the mean is chosen as a typical operator since it can share an information value interpretation [37].

### 2.2 Fire spread model

Fire spread models are commonly employed by fire behavior analysts to predict how fast a wildfire will spread [38]. An accurate fire spread model can help analysts develop community evacuation strategies and avoid potentially disastrous consequences. Since uncertainties are inevitable in model inputs, many scholars have conducted GSA for different fire spread models to enhance the understanding of fire behavior, such as GSA for the Rothermel model [39], the FARSITE model [40], the SPITFIRE model [41] and the Spark model [42].

In this work, we consider an empirical fire spread model established by Cheney et al [34] called Dry Eucalypt model, which is widely used for Australian eucalypt forests. The model has only four input variables, which enables a clear study on the effect of segmented characteristics on GSA results. In the model, the fire spread rate $R$ (m/h) can be expressed as follows:

$$R = \begin{cases} 30\varphi M_f, U \le 5 \\ \left[30 + 1.531(U-5)^{0.858} FHS_s^{0.93} \left(FHS_{ns}H_{ns}\right)^{0.637} \cdot 1.03\right]\varphi M_f, U > 5 \end{cases} \tag{13}$$

$$\varphi M_f = 18.35\left(2.76 + 0.124RH - 0.0187T\right)^{-1.495} \tag{14}$$

$$FHS_s = 3.39\left[1 - \exp\left(-0.03 \cdot 12FA\right)\right] \tag{15}$$

$$FHS_{ns} = 2.5\left[1 - \exp\left(-0.22 \cdot 12FA\right)\right] \tag{16}$$

$$H_{ns} = 23.33\left[1 - \exp\left(-0.025 \cdot 12FA\right)\right], \tag{17}$$

where $T$ is the air temperature; $RH$ denotes the relative humidity; $U$ is the average 10-m open wind speed; and $FA$ represents the fuel age. Detailed explanations of other parameters can be found in [34]. In this model, $R$ corresponds to different models depending on the indicator $U$, and the critical threshold of $U$ is 5km/hr.

### 2.3 Parameter settings

The normal or uniform distribution is always arranged for uncertainty analysis when no information is available [30]. In this analysis, the $i^{th}$ input variable is assumed as a random variable following the normal distribution with mean $\mu_i$ and standard deviation $\sigma_i$, $i = T$, $RH$, $U$, and $FA$. In practice, there may be correlations between variables, especially for the variable $T$ and $RH$. However, in order to investigate the effect of segmented characteristics on GSA results without other potential influence, all the variables are assumed to be independent of each other in this paper.

The uniqueness of uncertainty analysis for segmented models is the abrupt model change at the critical threshold. Since the indicator $U$ is uncertain, the overall output is jointly determined by both stages during the transition region. In order to explore the significant impact of segmented characteristics on GSA results, hereby, $\mu_U$ is treated as a varying parameter. The main purpose of varying $\mu_U$ is to (a) alter the output percentage of two stages during the transition region and examine the variation of GSA results and (b) examine the GSA results when the model is not in the transition region and not affected by the segmented characteristic.

Table 2 lists the corresponding distribution parameters. Since $U$ is a random variable following the normal distribution, we define the transition region according to the threshold (i.e., 5km/hr) and ± 3-sigma of $U$. For simplicity, we denote the phase where 2km/hr < $\mu_U \le$ 3.5km/hr as Stage 1, the phase where 6.5km/hr $\le \mu_U <$ 8km/hr as Stage 2, and the phase where 3.5km/hr < $\mu_U <$ 6.5km/hr as Stage 3. In other words, $\mu_U$ is a varying parameter with a variation range of 2-8km/hr and a variation interval of 0.1km/hr. It is clear that Stage 3 corresponds to the transition region and its overall output is composed of the output of two different stages.

Table 2 Basic random variables and the distribution parameters for the fire spread model.

| Random variable (unit) | Distribution | Mean | Standard deviation | Acceptable range |
|---|---|---|---|---|
| $T$ (°C) | Normal | 25 | 4 | 10 - 40 |
| $RH$ (%) | Normal | 20 | 2 | 14 - 26 |
| $U$ (km/hr) | Normal | $\mu_U$ | 0.5 | 0.5 – 9.5 |
| $FA$ (yr) | Normal | 4 | 0.8 | 1.5 – 6.5 |

*Note*: When the sampled value is out of the acceptable range, it will be resampled.

**Remark 1**. The applicable temperature range for the model is 10°C - 40°C [33], which is almost satisfied in the parameter settings. Besides, according to the experimental data for the dry eucalypt forest in the literature [34, 43], we choose the range of the *RH*, *U* and *FA* to be 14% - 26%, 0.5km/hr – 9.5km/hr and 1.5yr – 6.5yr, respectively. The range of variables chosen in this paper is essentially within the parameter range of the experimental data. When the sampled value is out of the given range, it will be resampled. It should be noted that the range and the distribution assigned to each input variable could easily be changed if required for other analysis [34].

**Remark 2**. It should be noted that GSA analysis has been carried out for the same fire spread model in [33]. However, the wind speed *U* follows a uniform distribution which is consistently greater than 5km/hr in their study. Since the fire spread behavior will have an abrupt change when *U* exceeds 5km/hr, they did not consider the effect of segmented characteristics (i.e., the abrupt model change) on the GSA results, which is the main contribution of this paper.

### 2.4 Estimation details

The numerical implementations of these four GSA indices are presented in the Appendix. For a single estimation, 4,000, 10,000, 5,000 and 2,000,000 samples are generated by Latin hypercube sampling [44] to calculate the Sobol index, mutual information, delta index and PAWN index, respectively. To ensure the robustness of the estimation results, the corresponding 95% confidence intervals are calculated. Specifically, the Jackknife resampling [45] is applied for the mutual information and the bootstrap resampling [46] is applied for the Sobol, delta and PAWN index. The number of Jackknife and bootstrap replication is set as 1000, which is determined empirically by the previous study [47].

**Remark 3**. This paper focuses on different importance rankings obtained by different GSA methods. Therefore, we ensure that the numerical implementations of the GSA indices are valid and the generated samples are sufficient to get convergence, without making the sample size of each method comparable. The convergence validation and the computation cost discussion can be found in the Appendix.

**Remark 4**. In the literature, the total effect $S_i^T$ is always chosen to be compared with the moment-independent index [48]. Hence, we choose $S_i^T$ to compare with the moment-independent approaches in this paper.

**Remark 5**. According to [45], resampling data with replacement produces duplicate data points. The *k*-nearest neighbor algorithm interprets these duplicates as detailed, high-information features, which causes an overestimation of mutual information in the bootstrapped samples. Therefore, jackknife resampling is used to estimate the confidence level of the mutual information rather than the bootstrap resampling, as suggested in [45].

## 3 Results and analysis

Following the numerical implementations in the Appendix and estimation details in Section 2.4, the importance rankings and corresponding 95% confidence intervals of different GSA methods under different $\mu_U$ are calculated. The results are illustrated in Fig. 1.

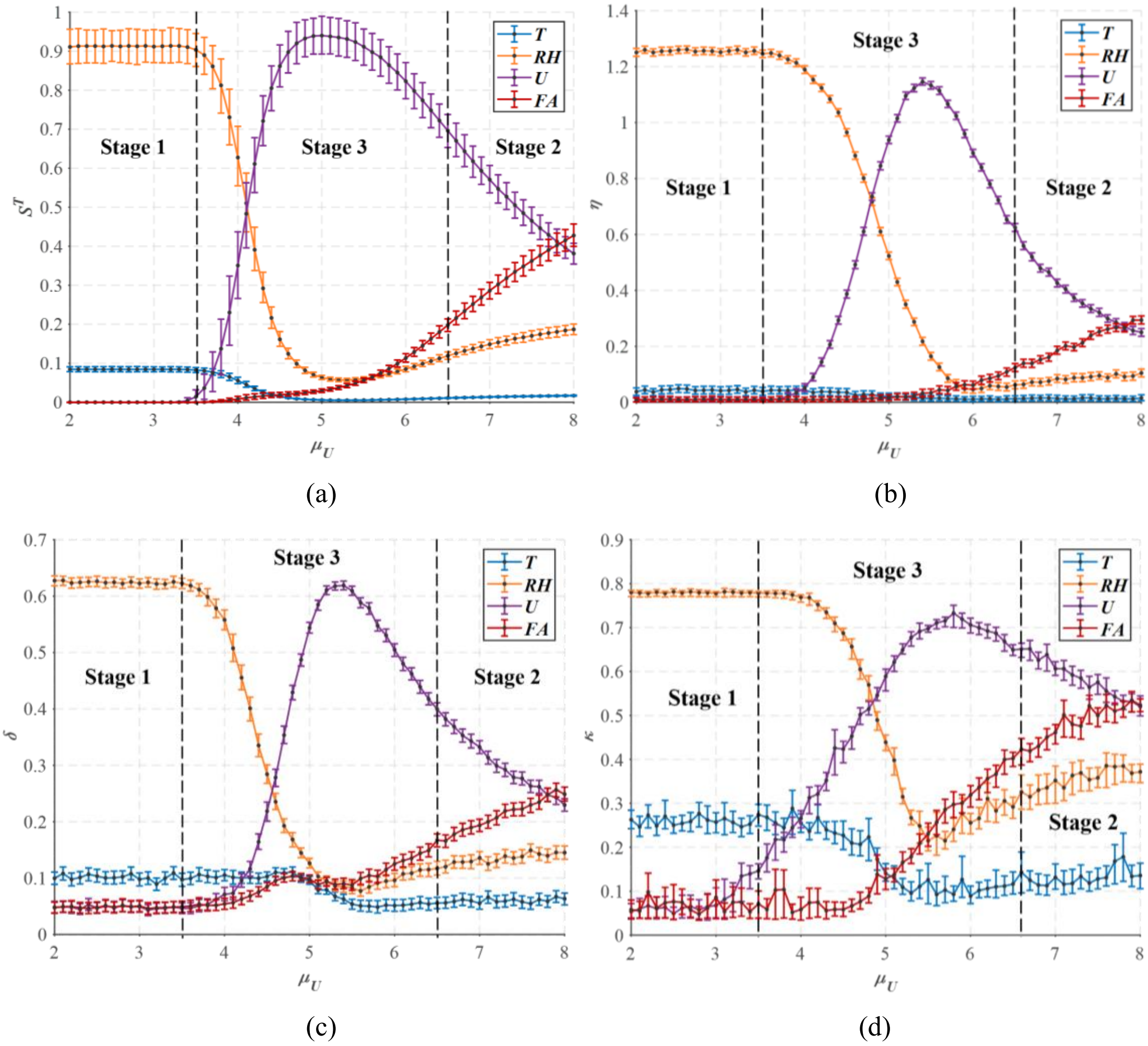

Fig. 1 GSA results under different $\mu_U$: (a) Sobol index (b) Mutual information (c) Delta index (d) PAWN index. The three stages are divided by the black dashed line. At Stage 1 or Stage 2, these four GSA indices yield the same importance rankings. Whereas at Stage 3, the importance rankings can be different.

As shown in Fig. 1, when the model is not in the transition region (i.e., at Stage 1 or Stage 2), these four GSA indices yield the same importance rankings. At Stage 1, *RH* has a decisive impact on the output uncertainty, followed by *T*, while *U* and *FA* have almost no influence. During Stage 2, in the beginning, *U* contributes most to the uncertainty of model output, followed by *FA*, while *T* and *RH* have little impact. As $\mu_U$ increases, the importance of *U* decreases and the importance of *FA* grows. When $\mu_U$ exceeds 7.8 km/hr, the importance rank of *U* and *FA* is exchanged.

Nonetheless, when the model is at Stage 3, the importance rankings obtained by these four GSA indices can be different. Specifically, for the variables *RH* and *U*, as $\mu_U$ increases, the importance of *RH* first decreases and then increases, while the importance of *U* first increases and then decreases. For the Sobol index, mutual information, delta index and PAWN index, the importance of *U* outranks *RH* when $\mu_U$ is up to 4.2, 4.8, 4.6 and 4.9 km/hr, respectively. Besides, *U* achieves the highest importance percentage when $\mu_U$ is up to 5, 5.4, 5.4 and 5.8 km/hr, respectively.

Fig. 1 illustrates the significant impact of segmented characteristics on the GSA results. In order to further reflect the differences in the importance rankings obtained by different GSA methods, we choose the GSA results when $\mu_U$=4.7km/hr as a specific case. Table 3 and Fig. 2 illustrate the importance rankings and corresponding 95% confidence intervals when $\mu_U$=4.7km/hr. From Table 3 and Fig. 2, we can see that the ranking results of these four GSA methods are not the same. For the mutual information and PAWN index, the importance ranking is *RH*>*U*>*T*>*FA*. For the delta index, the importance ranking is *U*>*RH*>*T*>*FA*. Whereas, for the Sobol index, the importance ranking is *U*>*RH*>*FA*>*T*.

Table 3 Global sensitivity results of the fire spread model.

| Variable | $S^T$ | 95% confidence interval | $\eta$ | 95% confidence interval |
|---|---|---|---|---|
| *T* | 0.0090 (4) | [0.0081,0.0100] | 0.0282 (3) | [0.0159,0.0413] |
| *RH* | 0.0987 (2) | [0.0889,0.1082] | 0.8011 (1) | [0.7868,0.8145] |
| *U* | 0.9203 (1) | [0.8680,0.9741] | 0.6106 (2) | [0.5973,0.6254] |
| *FA* | 0.0222 (3) | [0.0178,0.0272] | 0.0113 (4) | [0.0033,0.0210] |
| Variable | $\delta$ | 95% confidence interval | $\kappa$ | 95% confidence interval |
| *T* | 0.1091 (3) | [0.1016,0.1169] | 0.2066 (3) | [0.1805,0.2374] |
| *RH* | 0.1927 (2) | [0.1810,0.2050] | 0.6050 (1) | [0.5830,0.6269] |
| *U* | 0.3536 (1) | [0.3386,0.3694] | 0.5019 (2) | [0.4796,0.5174] |
| *FA* | 0.1003 (4) | [0.0929,0.1077] | 0.0766 (4) | [0.0593,0.0976] |

*Note*: The number in the parentheses represents the importance ranking of the variable for a specific GSA method.

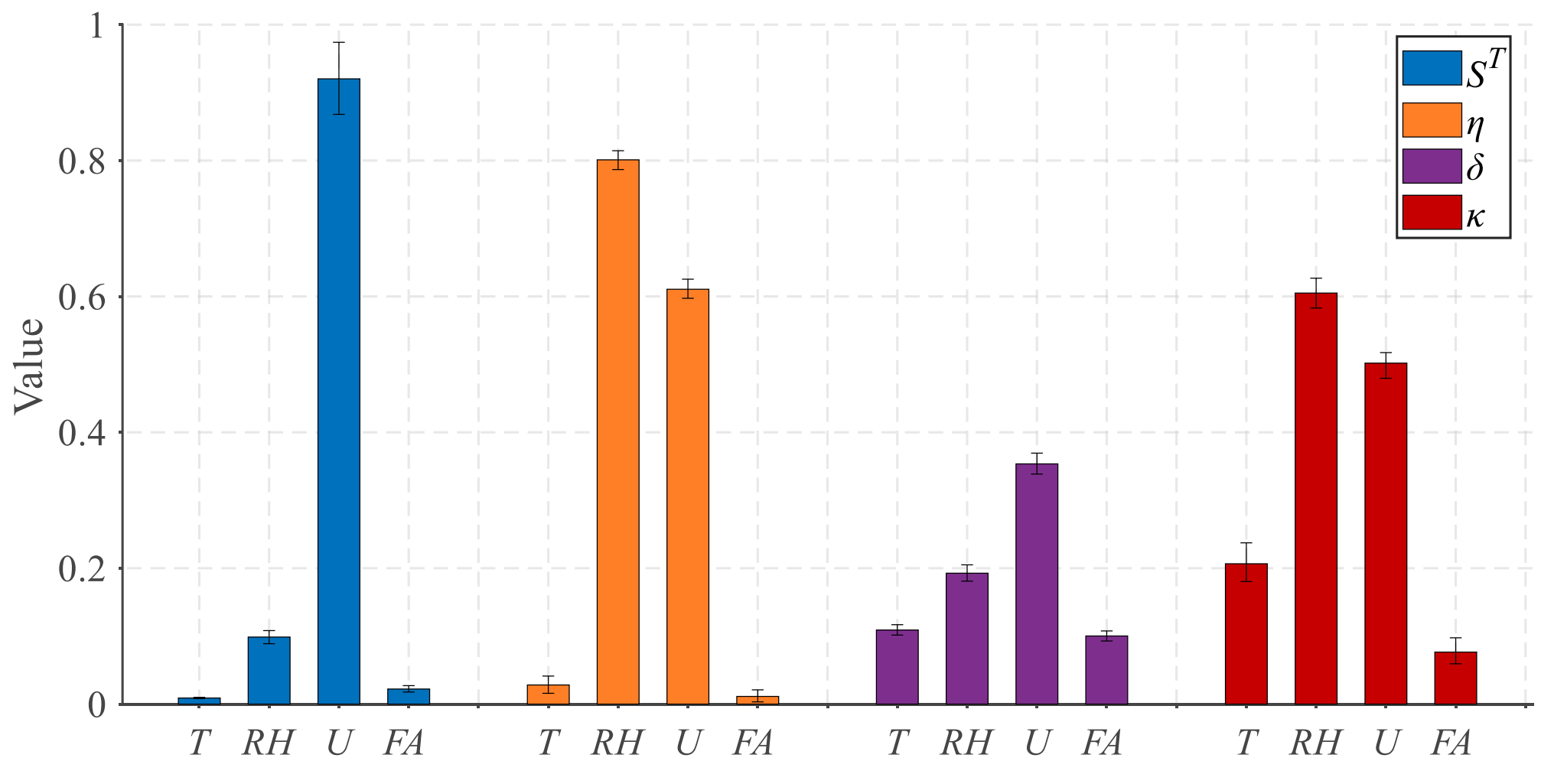

Fig. 2 Global sensitivity results of the fire spread model.

In order to provide an intuitive interpretation for the importance rankings of different GSA methods, we fix each variable to different values in turn and keep other parameter settings unchanged. Then, the corresponding conditional PDF and CDF are obtained and compared with the original PDF and CDF. Specifically, if the uncertainty of a variable has a great impact on the output uncertainty, then fixing it to different values will lead to significant variation in the output PDF and CDF. Conversely, if there is little effect, the output PDF and CDF will barely change. By this analysis, the properties of the each GSA measure can be further explored.

To make a reasonable comparison, since the input variables all follow the normal distribution in this analysis, we determine five fixed values based on the mean and standard deviation of each variable as follows: $\mu_i$-2$\sigma_i$, $\mu_i$-$\sigma_i$, $\mu_i$, $\mu_i$+$\sigma_i$ and $\mu_i$+2$\sigma_i$. Following the above settings, the conditional PDFs and CDFs of fixing *T*, *RH, U,* and *FA* are illustrated in Fig. 3- Fig. 6. In the following, we will analyze the importance rankings obtained by these four GSA methods in detail.

**Remark 6**. Choosing PDF and CDF for comparison is aimed to illustrate the reasons for the different importance rankings of four GSA methods since they can display the variation of output uncertainty visually and intuitively. This does not imply that the PDF or CDF based GSA method is more appropriate for uncertainty analysis in this model. How to choose appropriate GSA methods will be discussed later in Section 4.3.

**Remark 7**. The analytical approach used in this section is similar to the idea of the one-at-a-time method where the sensitivity is analyzed by varying one input variable at a time while keeping all others constant [49]. However, we adopt this idea here in order to interpret the importance rankings by different GSA methods, rather than quantify the importance of input variables directly like the one-at-a-time method.

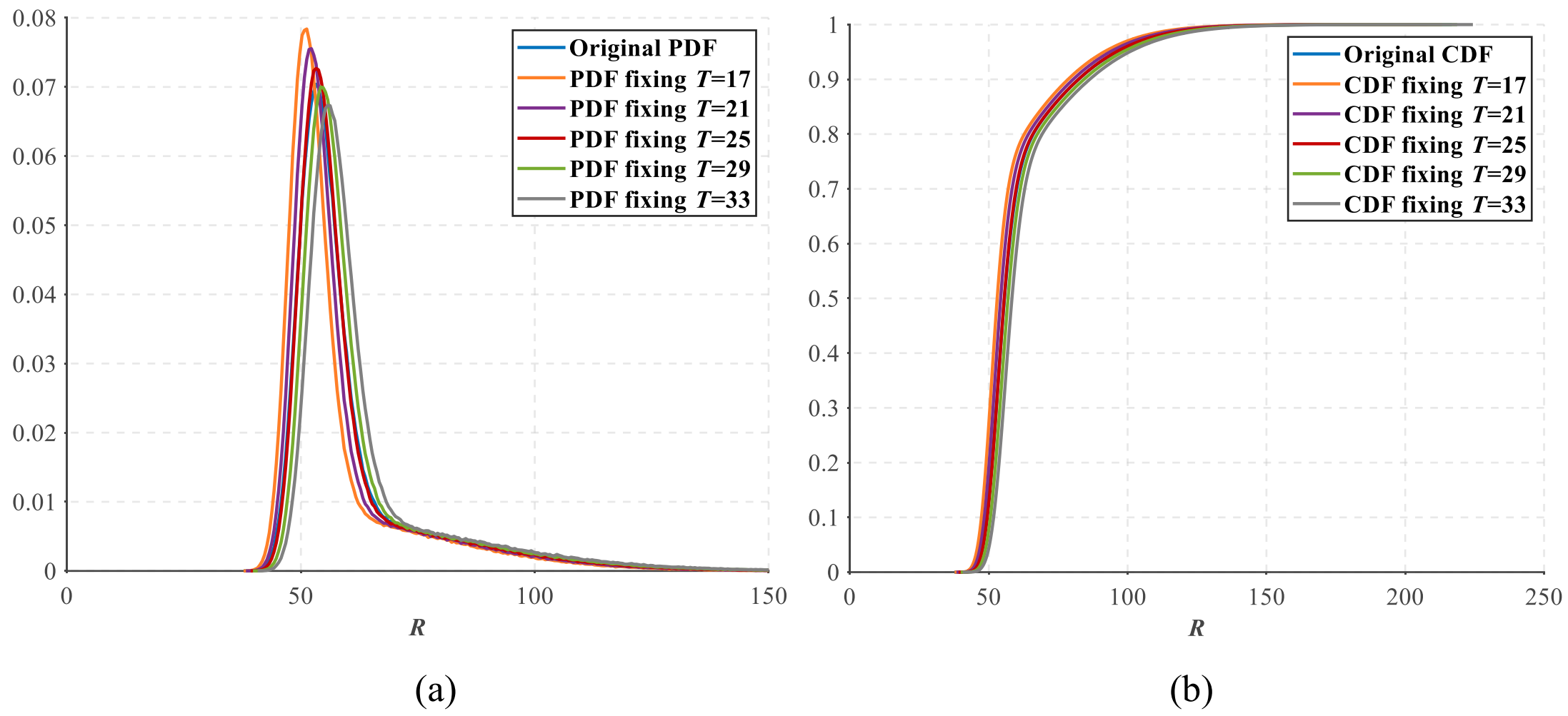

Fig. 3 PDF and CDF of *R* fixing *T* at different values. Fixing *T* affects the output range slightly.

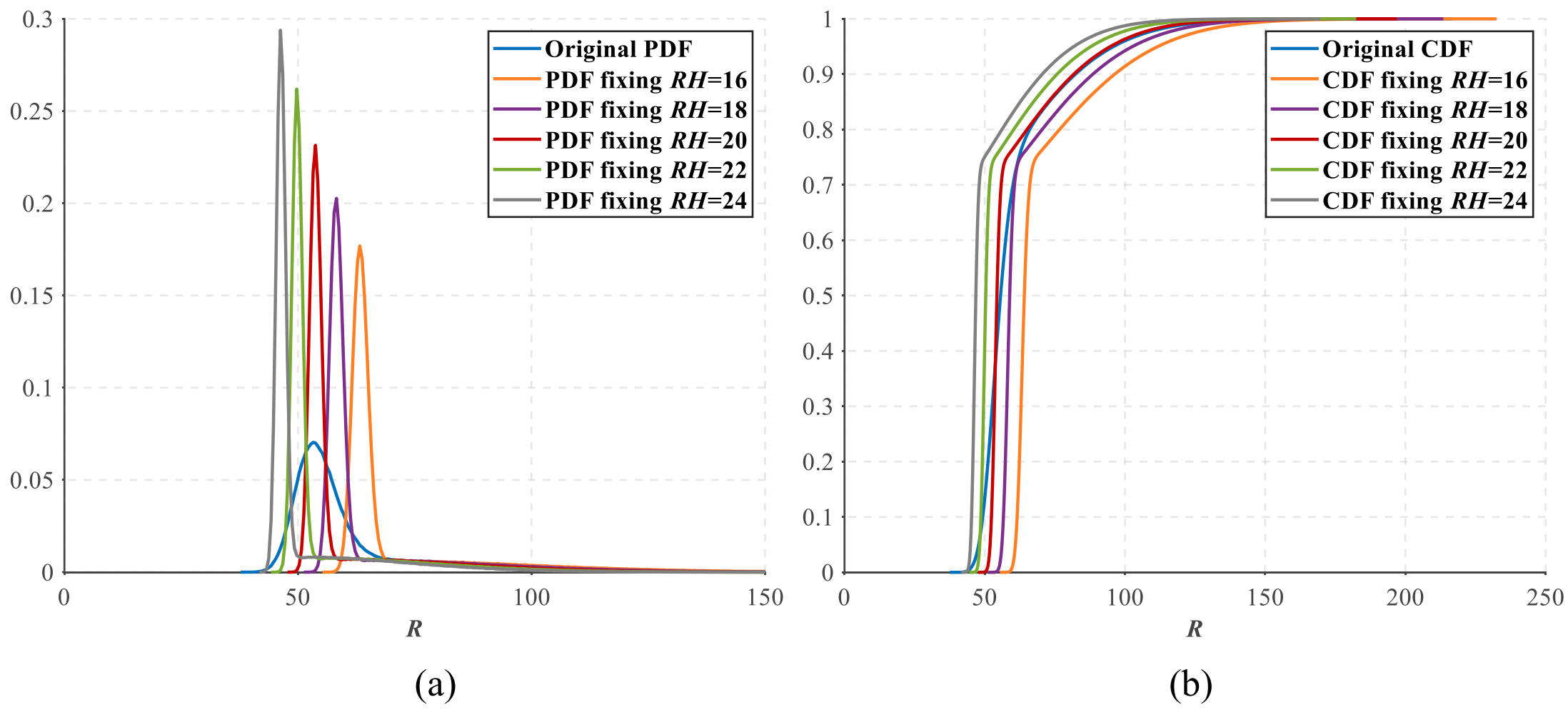

Fig. 4 PDF and CDF of $R$ fixing $RH$ at different values. Fixing $RH$ aggregates the output and affects the output range.

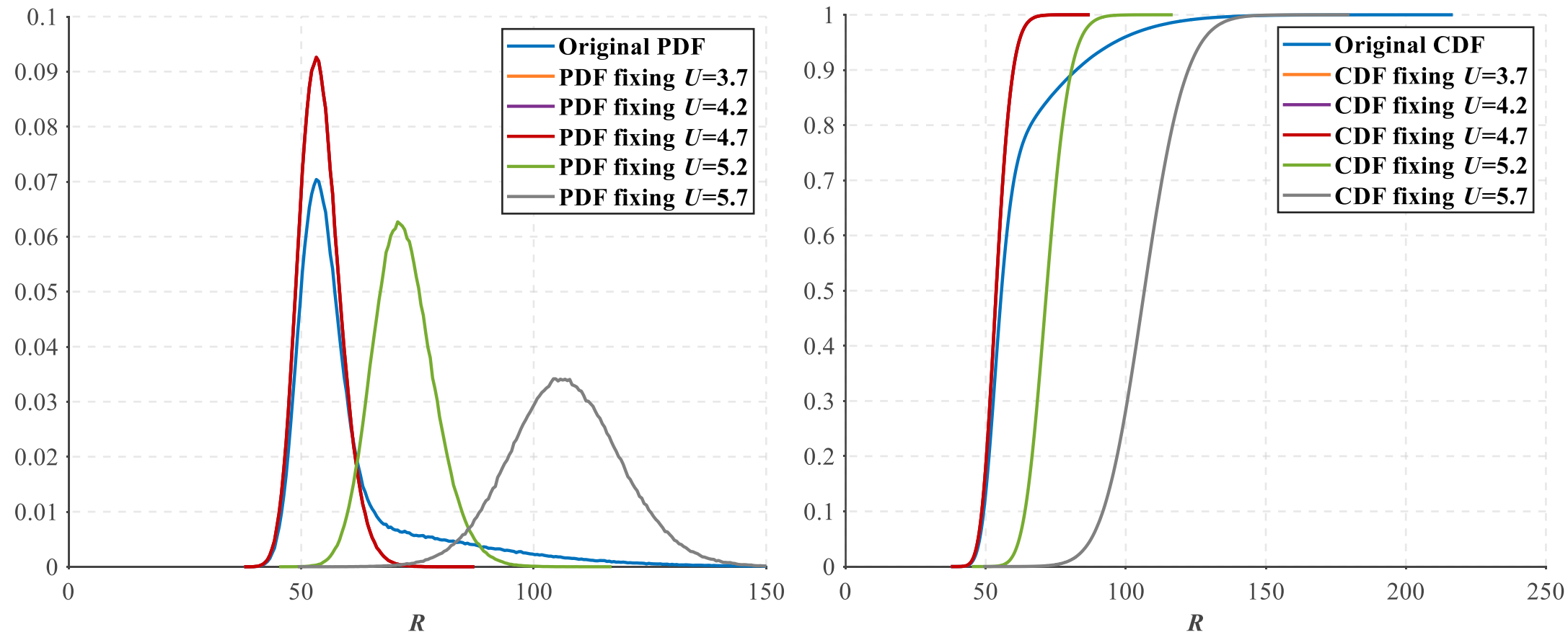

(a) (b)

Fig. 5 PDF and CDF of $R$ fixing $U$ at different values. Fixing $U$ eliminates the heavily-tailed characteristic and affects the output range significantly.

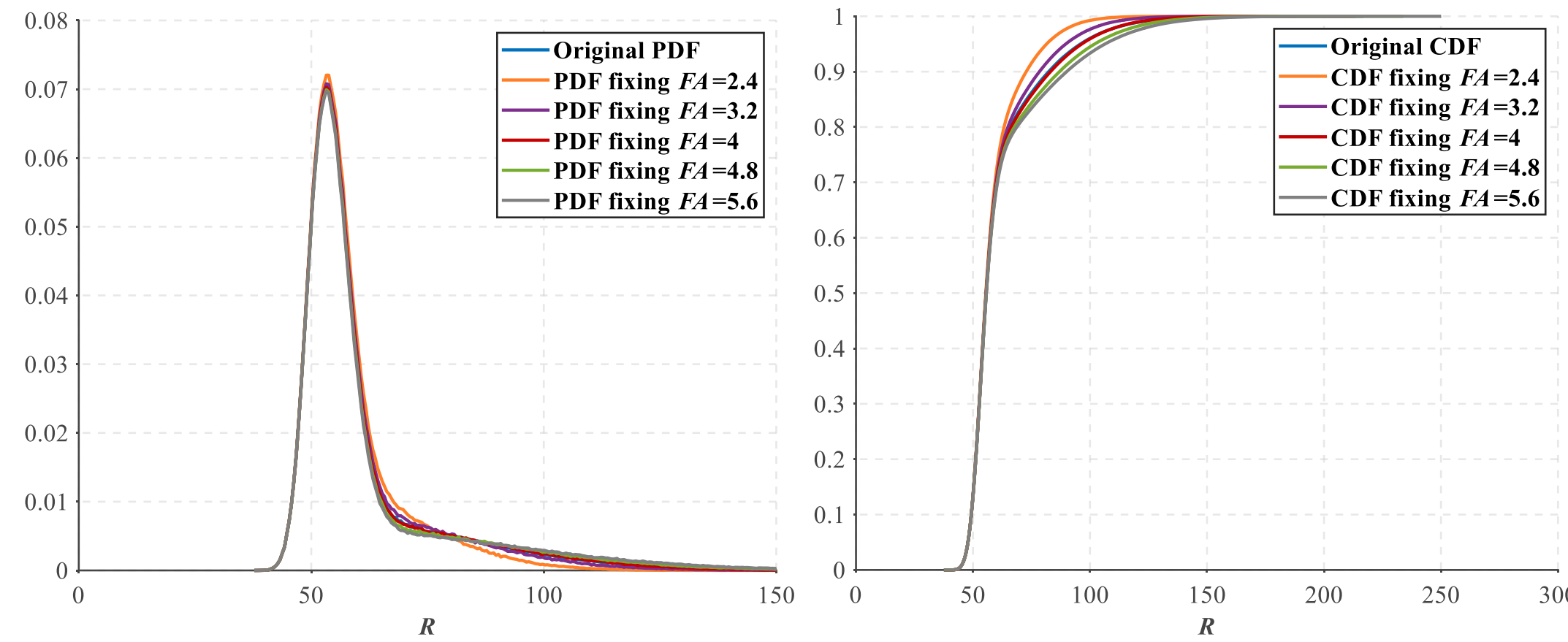

(a) (b)

Fig. 6 PDF and CDF of $R$ fixing $FA$ at different values. Fixing $FA$ affects the tail characteristics of the distribution.

**Sobol index**: Firstly, in order to understand the results of the Sobol index intuitively, we calculate the variance of the original output and the output when the variables are fixed at different values, respectively. The results are shown in Table 4. Then, as can be seen from Fig. 5, the original output PDF possesses a heavily-tailed characteristic, while fixing $U$ will eliminate this characteristic. This is because the heavily-tailed characteristic is caused by two points: the uncertainty of $U$ and the segmented characteristic of the fire spread model. When $U$ is fixed, the model is determined, thus eliminating the heavily-tailed characteristic. Since the heavily-tailed characteristic has a great impact on the variance, when it is eliminated, the output variance is greatly reduced. Therefore, the Sobol index supports that the variable $U$ is the most influential variable. From Fig. 4, it can be shown that although the output is more aggregated after fixing $RH$, the heavily-tailed characteristic still exists. Besides, when $RH$ is fixed to a small value, the heavily-tailed characteristic is more serious, making the output variance even larger than the original one. As a result, for the Sobol index, *RH* is less important than *U*. As for the variables *T* and *FA*, it is clear from Fig. 3 and Fig. 6 that fixing *FA* has a more significant effect on the heavily-tailed characteristic of the output than fixing *T*. Due to the great influence of the heavily-tailed characteristic on the variance, the Sobol index considers *FA* to be more important than *T*.

Table 4 Variance of the original output and the output when the variables are fixed at different values.

| Original variance | Variable | Fix variable at $\mu_i$-2$\sigma_i$ | Fix variable at $\mu_i$-$\sigma_i$ | Fix variable at $\mu_i$ | Fix variable at $\mu_i$+$\sigma_i$ | Fix variable at $\mu_i$+2$\sigma_i$ |
|---|---|---|---|---|---|---|
| 246.12 | *T* | 220.75 | 231.86 | 243.72 | 256.43 | 270.05 |
| | *RH* | 305.74 | 257.95 | 219.65 | 188.58 | 163.12 |
| | *U* | 19.69 | 19.69 | 19.69 | 42.91 | 144.59 |
| | *FA* | 112.21 | 178.43 | 246.70 | 310.79 | 367.48 |

**Mutual information**: Similarly, to make the results of mutual information intuitive, we calculate the entropy of the original output and the output when the variables are fixed at different values, respectively. The results are listed in Table 5. Then, as can be seen in Fig. 4 the output is more aggregated after fixing *RH*. Since the entropy pays attention to the aggregation of the entire PDF, fixing *RH* reduces the output entropy greatly. Therefore, the mutual information supports *RH* as the most important variable. From Fig. 5, it can be shown that fixing *U* contributes little to the aggregation level of the output PDF, leading to a limited reduction in the output entropy. In addition, when *U* is fixed to a large value, the output entropy even increases. Hence, for mutual information, *U* is less important than *RH*. As for the variables *T* and *FA*, it is clear from Fig. 3 and Fig. 6 that fixing *T* contributes more to the aggregation level of the output PDF than fixing *FA*. Since the aggregation level of the output has a large effect on the output entropy, the mutual information considers *T* to be more important than *FA*.

Table 5 Entropy of the original output and the output when the variables are fixed at different values.

| Original entropy | Variable | Fix variable at $\mu_i$-2$\sigma_i$ | Fix variable at $\mu_i$-$\sigma_i$ | Fix variable at $\mu_i$ | Fix variable at $\mu_i$+$\sigma_i$ | Fix variable at $\mu_i$+2$\sigma_i$ |
|---|---|---|---|---|---|---|
| 3.6475 | *T* | 3.5599 | 3.5919 | 3.6243 | 3.6572 | 3.6908 |
| | *RH* | 3.1275 | 3.0088 | 2.8969 | 2.7909 | 2.6907 |
| | *U* | 2.9043 | 2.9043 | 2.9043 | 3.3018 | 3.9162 |
| | *FA* | 3.4689 | 3.5810 | 3.6554 | 3.7066 | 3.7429 |

**Delta index**: It can be shown from Fig. 5 that fixing *U* causes a considerable change in the output range. Specifically, when *U* is fixed to a large value, the output range is completely different from the original one. Since the delta index evaluates the area difference between the original PDF and the conditional PDF, the change in output range greatly increases the area difference. Therefore, the delta index supports that *U* is the most important variable. For the variable *RH*, as can be seen in Fig. 4, although fixing *RH* also changes the output range, the change is limited and the heavily-tailed characteristic still exists. As a result, for the delta index, variable *RH* is less important than *U*. As for the variables *T* and *FA*, it is clear from Fig. 3 and Fig. 6 that fixing *T* has a greater effect on the output PDF than fixing *RH*, so the delta index considers *T* to be more important than *FA*.

**PAWN index**: From Fig. 5, although fixing *U* to a large value changes the output CDF significantly, when *U* is fixed to a value less than 5, it has limited change on the output CDF. On the other hand, as can be seen in Fig. 4, no matter whether *RH* is fixed at a large or small value, it will have a notable effect on the output CDF. Considering that the PAWN index chosen in this paper quantifies the mean value of the maximum absolute difference between CDFs across conditioning points, *U* is considered to be less important than *RH*. As for the variables *T* and *FA*, it is clear from Fig. 3 and Fig. 6 that fixing *T* changes the output CDF to a much greater extent than fixing *FA*. Therefore, the PAWN index supports that *T* is more important than *FA*.

From the above analysis, due to the segmented characteristics during the transition region, the combination of the output of two stages will lead to undesirable characteristics in the overall output distribution. Since GSA indices measure the uncertainty importance from different perspectives, when there are undesirable characteristics in the output distribution, it will lead to different importance rankings.

## 4 Discussions

In Section 3, we analyzed the importance rankings of different GSA methods for the segmented fire spread model. In this section, we firstly compare the results of the variance-based method and moment-independent methods in response to the prevalence of applying variance-based methods for segmented models in existing studies. Then, it is worth noting that both mutual information and delta index originate from PDFs but their importance ranking results are inconsistent. This interesting phenomenon is discussed in Section 4.2. Finally, how to choose an appropriate GSA method, which is of the most concern to analysts, is discussed in Section 4.3.

### 4.1 Comparison of variance-based and moment-independent methods

As mentioned in the introduction, most of the existing studies only use the variance-based method for conducting sensitivity analysis on segmented models. However, as can be seen from Table 3, there is a situation where none of the moment-independent methods obtain the same importance rankings as the variance-based method in the transition region of the segmented model. Moreover, if we denote the symbols $\sim$, $>$ and $\gg$ as differences in sensitivity between 0% and 10%, between 10% and 50%, and larger than 50%, respectively [37], then the Sobol index supports that $U \gg RH$, which is quite different from the importance rankings obtained by the moment-independent methods (the mutual information and the PAWN index support that $RH > U$, and the delta index supports that $U > RH$).

Such a result arises because the output PDF has a heavily-tailed characteristic, while variance only focuses on the second-order moment of the output and is very susceptible to the heavily-tailed characteristic. As seen in Fig. 5, fixing *U* can eliminate the heavily-tailed characteristic and greatly reduce the output variance. What's more, since the Sobol index is a normalized index, it will support that fixing *U* controls nearly all of the output uncertainty. Whereas, for the moment-independent methods, since they focus on the whole distribution and are finitely influenced by the heavily-tailed characteristic, they will not get radical results. Therefore, if the variance-based method is adopted without considering the actual requirements, a misleading importance ranking may be obtained, leading to suboptimal or even wrong decision making.

### 4.2 Comparison of mutual information and delta index

Mutual information and delta index are both GSA indices based on the output PDF. However, as can be seen from Table 3, these two methods obtain different importance rankings for the segmented fire spread model. Specifically, mutual information supports *RH* to be more important than *U*, while the delta index supports that *U* is the most influential variable. The same situation occurs in [26] where these two GSA indices give different importance rankings. In essence, the reason for the different rankings lies in the different focus of the two indices. Mutual information quantifies the difference in Shannon entropy between the original output and the conditional output. Shannon entropy is an uncertainty measure for random variables and is not related to the change in the output range. On the other hand, the delta index quantifies the area difference between the original PDF and the conditional one. It is a distance-based uncertainty importance measure that cannot quantify the uncertainty of random variables. Meanwhile, it is highly influenced by the variation of the output range.

As for the segmented fire spread model, the output range of the two stages has considerable differences. Since the model stage is determined by the value of the indicator *U*, different *U* can greatly affect the output range, as illustrated in Fig. 5. When analyzing the contribution of the uncertainty of *U*, the mutual information compares the output entropy without considering the change in the output range, so it supports that *U* is less important than *RH*. In contrast, the definition of the delta index makes it subject to the change in the output range, so it supports that *U* is the most influential variable. Therefore, these two GSA methods get different importance rankings.

### 4.3 How to choose an appropriate GSA method for the segmented model

Through the results and analysis in Section 3, segmented characteristics indeed lead to different importance rankings for different GSA metrics indices. Then, an intuitive and essential question is which GSA index should be chosen as the basis for decision making. As argued in [37], the choice of a GSA method should be tied to the analyst's anticipated audience or specific requirements. For example, if analysts are concerned with the change in the output distribution, then the moment-independent approaches are preferred. On the other hand, if the desired report includes some measure of central tendency, the variance-based method may be more appropriate [37].

In terms of the segmented fire spread model during the transition region, the impact of the segmented characteristic on the GSA indices stems from the following points:

- The output uncertainty varies at different stages. Then, due to the uncertainty in the indicator *U*, the overall output is jointly determined by both stages, and the combination of the output of the two stages leads to heavily-tailed characteristics in the overall output PDF.
- The importance of different variables varies at different stages. Therefore, the heavily-tailed characteristic of the output PDF may be more significant after fixing certain variables (e.g., fixing *RH*).
- Fixing the indicator *U* will eliminate the heavily-tailed characteristic of the output PDF and significantly affect the output range.

Due to the above particularities of the segmented model, different GSA indices have different features when applied to the segmented model, which are summarized as follows:

- For the Sobol index, as discussed in Section 4.1, a radical importance ranking is obtained, and fixing the indicator *U* controls nearly all of the output uncertainty. At this point, variance-based methods may not be recommended as the basis for decision making.
- For the mutual information, it evaluates the variation in the aggregation level of output without being affected by the undesirable characteristics arising from segmented characteristics. If analysts are concerned about the variation of the aggregation level of output PDF, mutual information should be preferred.
- For the delta index, as discussed in Section 4.2, the variation of the output range is concerned, and the contribution of the indicator *U* enlarges. If the variation of the output range is of concern for analysts, the delta index should be preferred.
- For the PAWN index chosen in this paper, it obtains a robust importance ranking result, and the influence of the undesirable characteristics arising from segmented characteristics is relatively small. If analysts are concerned about the variation of output CDF, the PAWN index should be preferred.

Although the above findings do not uniquely identify the appropriate GSA approach under all circumstances, they do narrow the field and provide reasonable justification for the final choice of method. In addition, the above analysis can enhance our understanding of different GSA methods and provide guidance in choosing appropriate GSA methods for other segmented models.

## 5 Conclusion

In this paper, we focus on the implementation of GSA on a special class of mathematical models, i.e. segmented models. Sobol index, mutual information, delta index and PAWN index are used to carry out GSA for a segmented fire spread model called Dry Eucalypt model in the transition region. Some conclusions could be drawn as follows:

- We demonstrate that the inconsistent GSA results stem from the segmented characteristic. During the transition region, these four GSA methods get different importance rankings. In contrast, when not in the transition region, these four GSA methods yield the same GSA results regardless of the value of $\mu_U$.
- The variance-based approach may not be suitable for dealing with the segmented model during the transition region, as it will yield a radical importance ranking result, i.e., fixing the indicator controls nearly all of the output uncertainty.
- For moment-independent methods, the mutual information is concerned with the aggregation level of the output PDF, the delta index takes into account the variation of the output range, and the PAWN index is concerned with the variation of output CDF. All these GSA indices can be used as the basis for decision making when dealing with segmented models, depending on the specific requirements of analysts.
- For the Dry Eucalypt model, during its transition region, *U* is always an influential variable and *T* always plays a minor role; when close to the Stage 1, *RH* has significant impact and *FA* has little impact; when close to the Stage 2, *FA* is more influential than *RH*.

The work in this paper clarifies the necessity for applying appropriate indices when performing GSA on a segmented fire spread model, especially during the transition region. Analysts should value the uniqueness of the segmented model and consider the results of different GSA indices according to their practical purpose, so as to provide credible information for decision making. All of our source codes are publicly available at https://github.com/dirge1/GSA_segmented.

Although this work carries out a detailed analysis for the segmented fire spread model, the generalization of the findings needs to be verified on a more extensive set of segmented models. In the supplementary material, the GSA results of another segmented model during the transition region are compared to illustrate the generalizability of the findings in this paper. Furthermore, we will work on providing a mathematical basis for the impact of segmented characteristics on different GSA approaches in future research, not limited to case studies.

## Acknowledgements

This work was supported by the National Natural Science Foundation of China [grant number 51775020], the Science Challenge Project [grant number. TZ2018007], the National Natural Science Foundation of China [grant numbers 62073009].

## Appendix

### A. Numerical implementations for GSA methods

The implementation methods of these four GSA indices are summarized in Table 6 and expatiated in the following.

**Sobol index**: Since computing variance using analytical integrals is always challenging, Monte Carlo integration is applied to get a numerical solution [50]. Following that, $S_i$ and $S_i^T$ can be rewritten as

$$\hat{S}_i = \frac{1}{N \cdot \hat{V}(Y)} \sum_{k=1}^{N} f(\boldsymbol{x}_k) \cdot \left[ f(\boldsymbol{x}_k^i) - f(\boldsymbol{x}_k') \right], \tag{18}$$

$$\hat{S}_i^T = \frac{1}{2N \cdot \hat{V}(Y)} \sum_{k=1}^{N} \left[ f(\boldsymbol{x}_k^i) - f(\boldsymbol{x}_k') \right]^2, \tag{19}$$

where

$$\boldsymbol{x}_k = (x_{k1}, x_{k2}, \cdots, x_{kn}) \tag{20}$$

$$\boldsymbol{x}_k' = (x_{k1}', x_{k2}', \cdots, x_{kn}') \tag{21}$$

$$\boldsymbol{x}_k^i = (x_{k1}', x_{k2}', \cdots, x_{ki}, x_{k(i+1)}', \cdots, x_{kn}') \tag{22}$$

are three sample vectors obtained from the $k^{\text{th}}$ row of two independent sample matrices (i.e., $\boldsymbol{x}$ and $\boldsymbol{x}'$) for $\boldsymbol{X}$; $N$ is the number of rows of the sample matrix; and $\hat{V}(\cdot)$ is the variance estimation. Notably, the Sobol index is called moment-dependent because the estimation of variance is based on the second-order moment given as:

$$\hat{V}(Y) = \frac{1}{N-1} \sum_{i=1}^{N} (y_i - \bar{y})^2, \tag{23}$$

where

$$\bar{y} = \frac{1}{N}\sum_{i=1}^{N} y_i. \tag{24}$$

**Mutual information**: The $k$-nearest neighbor (KNN) algorithm is applied to calculate the mutual information [51]. Compared to density estimation techniques, the KNN method utilized the distribution of $k^{\text{th}}$ nearest neighbor distance to represent the real density, thus avoiding the challenge of estimating the joint density directly. Besides, this computation technique is fast and efficient, and only requires a single hyperparameter $k$ which has a relatively small effect on the results. In this paper, $k$ is set as 3.

**Delta index**: We adopt an efficient estimation method introduced in [52]. Kernel-density estimation is applied to estimate $f_Y(y)$ and $f(y|x_i)$. The method can not only give a robust result from given data, but also achieve a notable reduction in computational burden, making the estimation cost independent of the number of input variables.

**PAWN index**: The approximation method introduced in [53] is employed. This method can be performed directly from a generic dataset without tailored sampling. Besides, it only requires a single hyperparameter $n$ corresponding to the conditioning intervals and is robust against the choice of $n$. In this paper, $n$ is set as 10.

Table 6 Implementation methods and tools for the four GSA indices.

| GSA index | Implementation method | Implementation tool |
|---|---|---|
| Sobol index | Monte Carlo integration [50] | Matlab: authors' code |
| Mutual information | KNN algorithm [51] | Python: ennemi [54] |
| Delta index | Kernel-density estimation [52] | Python: SALib [55] |
| PAWN index | Generic sampling [53] | Matlab: SAFE [56] |

B. Convergence validation and computation costs

In this section, we validate the estimation convergence in Table 3 through a specific case that $\mu_U$=4.7km/hr. Convergence analysis is conducted by examining the stability of the importance rankings with increasing sample size. For each sample size, we conduct 100 replications of the estimation and take the mean value. The results are illustrated in Fig. 7. It can be seen that the importance rankings are stable with increasing sample size, which demonstrates that the estimation of GSA indices using the sample size chosen in this paper are convergent.

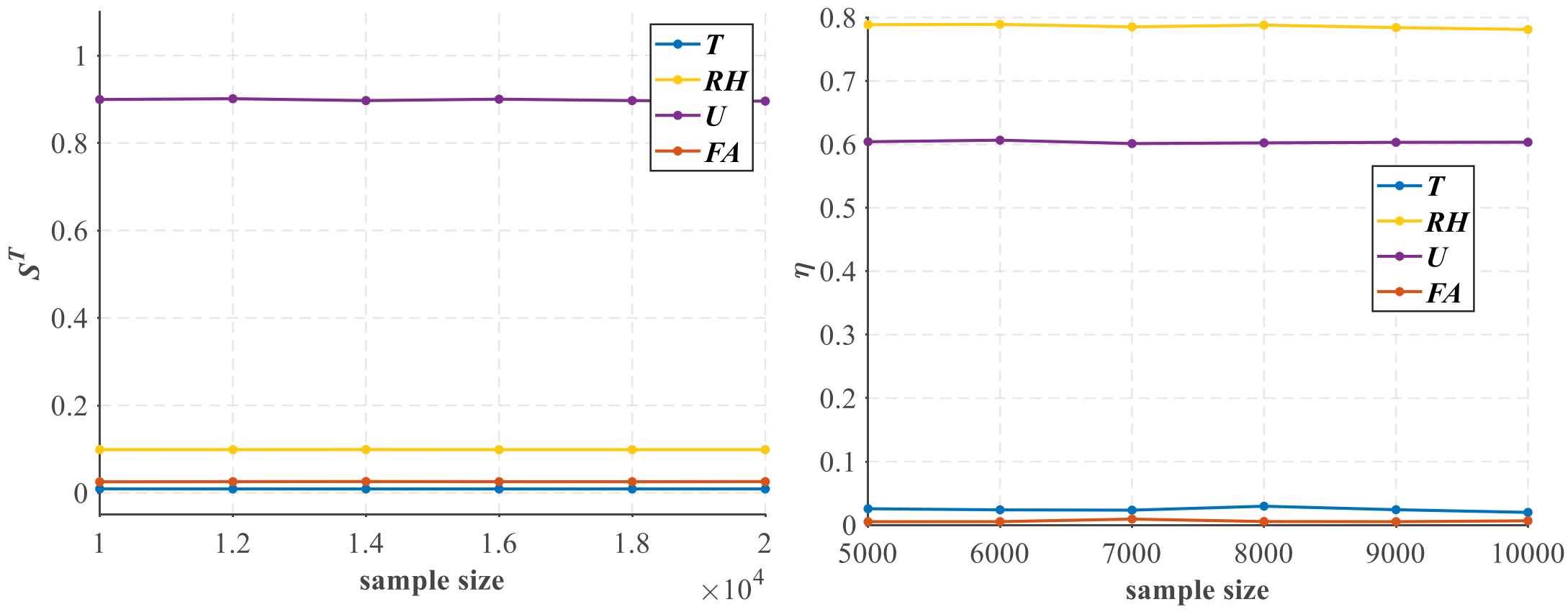

(a) (b)

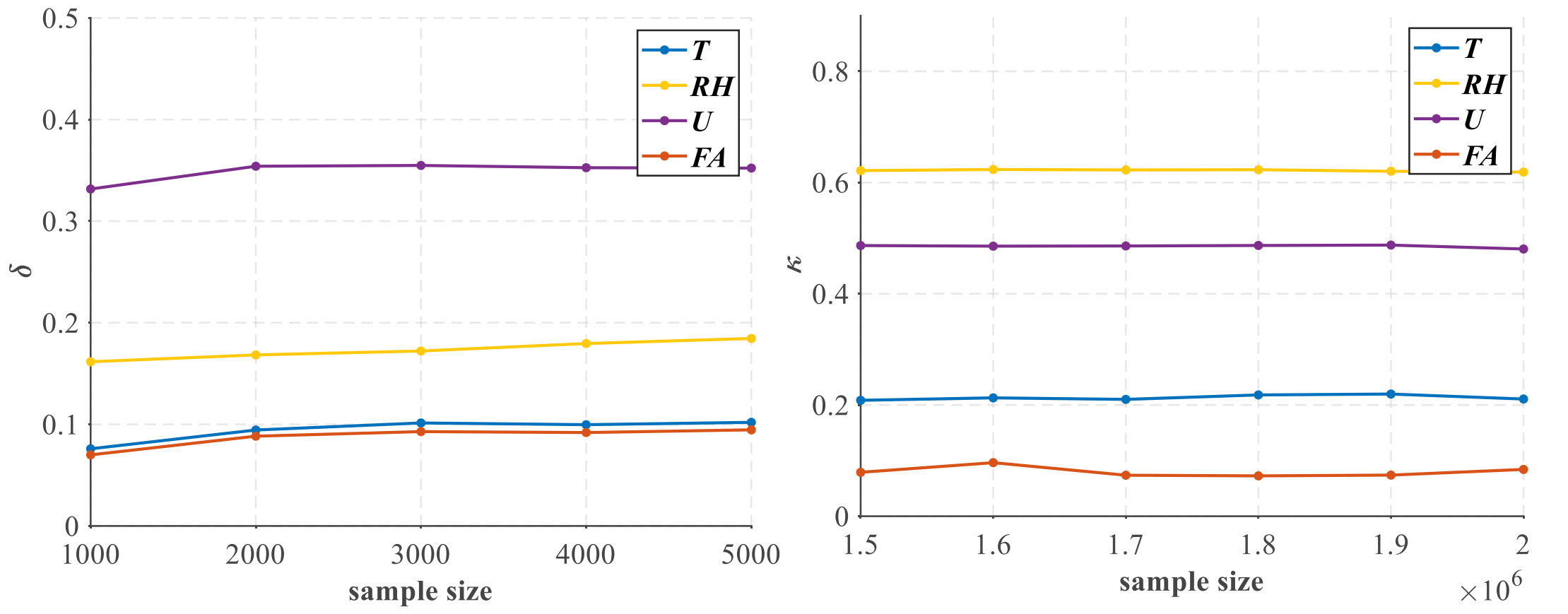

(c) (d)

Fig. 7 Importance rankings of different GSA methods under different sample size when $\mu_U$=4.7km/hr: (a) Sobol index (b) Mutual information (c) Delta index (d) PAWN index. The results indicate that the estimations obtained using the sample size chosen in this paper are convergent.

Besides, the computation time of each index under different sample sizes is shown in Fig. 8. It can be viewed that under the same sample size, calculating the delta index is the most time-consuming, followed by mutual information and PAWN index, while computing the Sobol index requires the least computational resources.

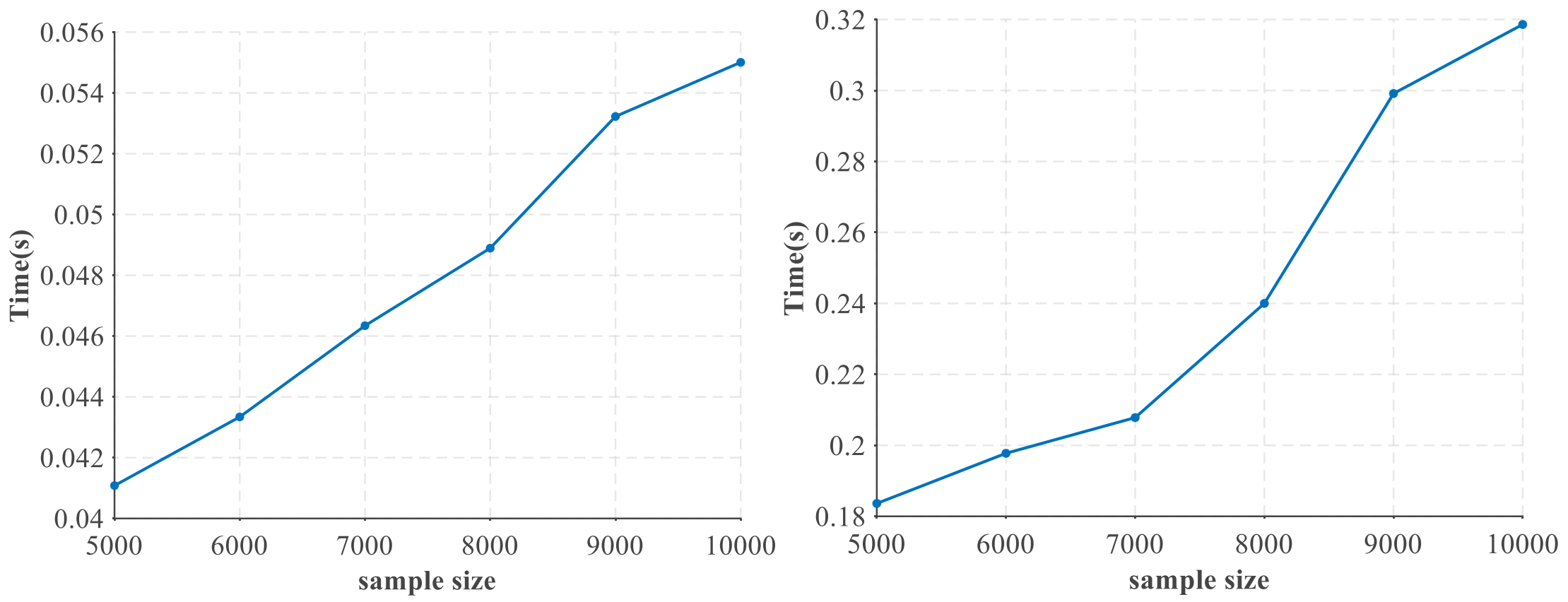

(a) (b)

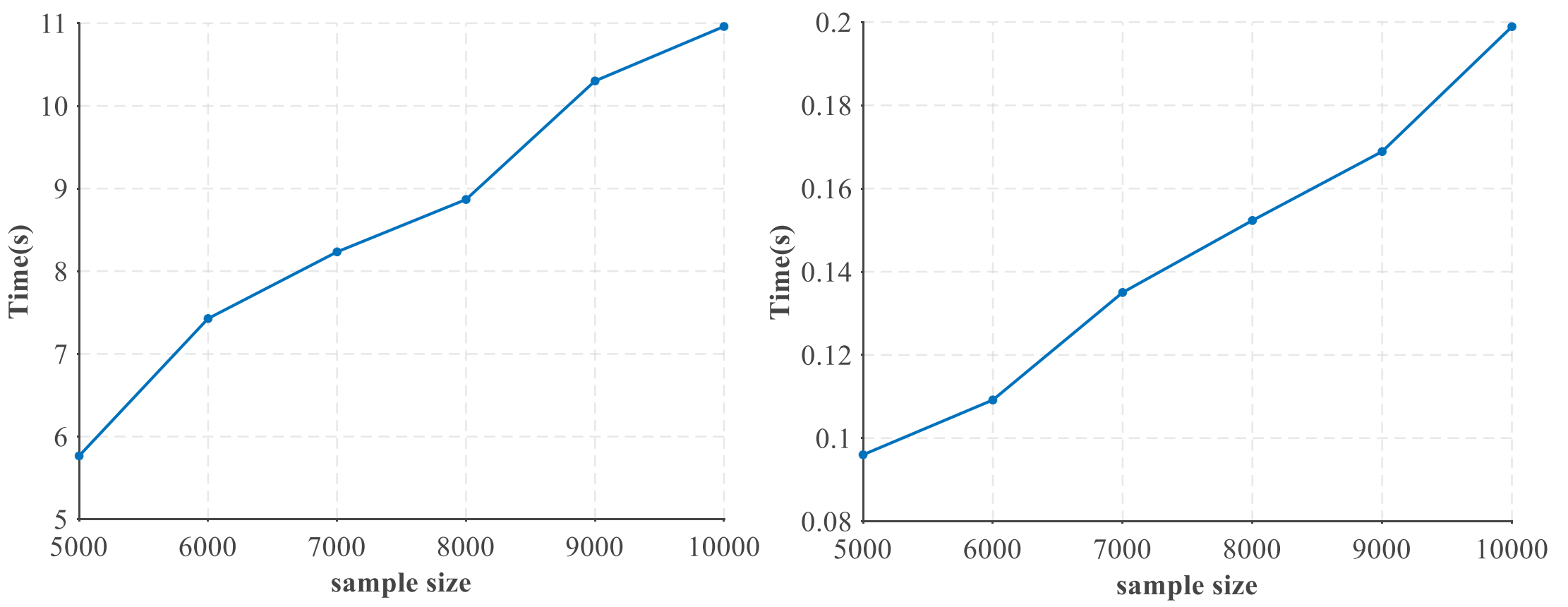

(c) (d)

Fig. 8 Computation time of different GSA methods under different sample size: (a) Sobol index (b) Mutual information (c) Delta index (d) PAWN index.